\documentclass[reprint,amsmath,amssymb,aps,superscriptaddress,nofootinbib,prd]{revtex4-2}
\usepackage{graphicx}
\usepackage{dcolumn}
\usepackage{bm}
\usepackage[utf8]{inputenc}
\usepackage{amsmath, amsthm, amsfonts, amssymb}
\usepackage[svgnames]{xcolor}
\definecolor{DarkRed}{RGB}{179, 27, 27}
\colorlet{color1}{NavyBlue}
\usepackage[colorlinks=true,allcolors=DarkRed]{hyperref}
\usepackage{bm}
\usepackage{physics}
\usepackage{dsfont}
\usepackage{graphicx}
\usepackage{cleveref}
\usepackage{xfrac}
\usepackage{mathrsfs}
\usepackage{comment}

\def\apjl{{ApJL}}

\def\mnras{{MNRAS}}

\def\prd{{Physical Review D}}
\def\prl{{Phys. Rev. Lett.}}

\def\04a{{2004 a}}
\def\04b{{2004 b}}

\newcommand{\LEM}{\mathcal{L}_{\rm EM}}

\begin{document}

\title{General Framework for the Spontaneous Scalarization of Regular Black Holes}

\author{Ernesto Contreras}
\affiliation{Departamento de F\'{\i}sica, Universidad de Alicante, Campus de San Vicente del Raspeig, E-03690 Alicante, Spain}
\author{Mikaela Carrasco-Hidalgo}
\affiliation{School of Physical Sciences and Nanotechnology, Yachay Tech University,  Urcuqu\'i, 100119, Ecuador}
\author{Pedro Bargue\~no}
\affiliation{Departamento de F\'{\i}sica, Universidad de Alicante, Campus de San Vicente del Raspeig, E-03690 Alicante, Spain}

\author{Arthur G. Suvorov}\email{arthur.suvorov@tat.uni-tuebingen.de}
\affiliation{Theoretical Astrophysics, Institute for Astronomy and Astrophysics, University of T\"{u}bingen, 72076 T\"{u}bingen, Germany}
\affiliation{Departamento de F\'{\i}sica, Universidad de Alicante, Campus de San Vicente del Raspeig, E-03690 Alicante, Spain}

\begin{abstract}
We investigate the spontaneous scalarization of generic, static, and spherically symmetric regular black holes supported by nonlinear electrodynamics. Starting from an arbitrary seed metric, we employ the P–dual formalism to reconstruct the electromagnetic sector and subsequently couple a real scalar field nonminimally. As a worked example, we apply the framework to the regular Balart–Vagenas black hole, showing that scalarized and scalar–free branches can coexist in a region where the scalarized configurations are entropically preferred. We further assess possible observational imprints, finding percent-level deviations in both the shadow size and the fundamental scalar quasi–normal modes ($< 10\%$ for small charge-to-mass ratios), indicating that current electromagnetic and gravitational–wave observations do not rule out these solutions. Our construction thus provides a general route to explore scalarization on top of nonlinear–electrodynamics–supported spacetimes, extending beyond specific Reissner–Nordström–like cases.
\end{abstract}

\maketitle

\section{Introduction} \label{sec:intro}

Black holes (BHs) are a core prediction of the theory of general relativity (GR). In electrovacuum, the Reissner-Nordstr{\"o}m (RN) and Kerr-Newman solutions uniquely describe these objects and yet host singularities \cite{penrose65,heus96}. This uncomfortable fact has led to investigations of how ``hair'' may be introduced into the action to regularize them (see Ref.~\cite{lan23} for a review). One such path involves nonlinear electrodynamics (NLED) \cite{Bronnikov:2000vy,Bronnikov:2017sgg,rin21,bronn23}: although Coulombic repulsion is not strong enough to prevent gravitational collapse, if the electromagnetic sector is augmented in the high-energy limit the attractive nature of gravity can be counteracted at some characteristic length-scale in such a way that collapse halts before a singularity forms for arbitrarily small charge. 

With the above in mind, several studies have been undertaken to essentially approach the inverse problem: how must the electromagnetic Lagrangian behave in order to produce a regular solution? A powerful tool in this direction is the so-called ``P-dual formalism'' \cite{pell69,sal87}. By making use of Legendre transforms, the dynamics in the electromagnetic sector can be rewritten such that a simple relation between the spacetime mass function and the (monopolar) electric and/or magnetic fields arises \cite{Ayon-Beato:1998hmi,Ayon-Beato:1999kuh,Dymnikova:2015hka,Carrasco:2024} (see also Ref.~\cite{suv21} for scalar-tensor constructions). This equation, and its generalisations to stationary BHs, can then be solved given some seed metric to engineer an appropriate action. Provided certain physical conditions are met (e.g. the weak-energy condition), NLED couplings can address the singularity problem and alter BH structure (though see also Refs.~\cite{dym92,Bronnikov:2000vy,boku22} regarding no-go theorems and ways to evade them). 

Aside from high-energy corrections to the electromagnetic sector, GR itself may be modified in the ultraviolet limit. One of the simpler, physically-motivated ways involves scalar couplings where fundamental ``constants'' are promoted to dynamical fields, as in the classical Brans-Dicke theory \cite{esp01}. In cases where the scalar field is non-minimally coupled to the electromagnetic and/or gravitational sectors, a phenomenon known as spontaneous scalarization can occur \cite{Damour:1993hw}. The underlying mechanism involves a turnover in the sign of the effective mass appearing within the Klein-Gordon equation describing the scalar dynamics. Such a transition implies the existence of exponentially growing (rather than damped) modes, leading to the growth of a scalar field that otherwise lies dormant in regions of low mass-energy thereby avoiding tensions with Solar system and other experiments (see Ref.~\cite{doneva24} for a review). In such theories, there may therefore exist multiple solution branches with rich dynamics that could, in principle, provide smoking-gun signatures for beyond-GR physics.

In this work, we study the spontaneous scalarization that arises from the coupling between a regular BH solution supported by NLED and a nonminimally coupled scalar field. The method is general and works for an arbitrary (static) seed, and therefore generalizes studies that have considered a similar problem but in the restricted contexts of the RN \cite{Herdeiro:2018wub}, power-Maxwell \cite{Carrasco:2024}, Bardeen \cite{zhang24,huang25}, or Euler-Heisenberg \cite{zhang25} BHs. After providing a general recipe and discussing numerical techniques, the method is demonstrated for a simple case with a \emph{regular} BH (namely the Balart-Vagenas solution \cite{bv14}). We explore some astrophysical manifestations for the new solutions found here, for example as concerns shadows and quasi-normal modes (QNMs).

This paper is organized as follows. Section~\ref{sec:spacetime} introduces the spacetimes we consider, together with the P-dual formalism and its application to the inverse problem (Sec.~\ref{sec:inverse}). Scalarization and the numerical methods we use are given in Section~\ref{sec:scalarization}. A worked example of the methodology is presented in Section~\ref{sec:examplesol}, with astrophysical connections explored in Section~\ref{sec:astro}. Some closing discussion is then provided in Section~\ref{sec:conclusions}.

\section{Black hole spacetime} \label{sec:spacetime}

We consider a static and spherically-symmetric line element in Schwarzschild-like coordinates $\{t,r,\theta,\varphi\}$ in the form 
\begin{equation}\label{eq:metric}
    \text{d}s^2=-N(r)e^{-2\delta(r)}\text{d}t^2+\frac{\text{d}r^2}{N(r)}+r^2\text{d}\Omega^2\ ,
\end{equation}
for some functions $N$ and $\delta$. 

\subsection{Inverse problem and the P-dual formalism} \label{sec:inverse}

In traditional approaches, one fixes a theory of gravity and solves for the metric potentials subject to some appropriate boundary conditions. By contrast, one can make use of the P-dual formalism to instead build a theory around \emph{given} choices of potentials \cite{pell69,sal87}. Provided care is taken to ensure a Maxwellian limit and other physical properties (e.g., weak- and other energy conditions \cite{cur14}), this approach allows for the exploration of a wide class of regular BH spacetimes (see, e.g., Refs.~\cite{Ayon-Beato:1998hmi,Ayon-Beato:1999kuh,Dymnikova:2015hka}).

Consider, at first, GR minimally-coupled to some theory\footnote{One might also consider the electromagnetic sector to depend on the Hodge-dual invariant, $G = \tfrac{1}{4}F^{\star}_{\mu \nu}F^{\mu \nu}$. As we do not consider dyonic cases in this paper and work with static spacetimes, this complication can be ignored without loss of generality.} of NLED via
\begin{equation} \label{eq:bareaction}
        S=\int \text{d}^4x \sqrt{-g} \left(R-\LEM \right),
\end{equation}
where $\LEM \equiv \LEM(F)$ is a function of the electromagnetic invariant, $F = \tfrac{1}{4}F_{\mu \nu}F^{\mu \nu}$ for Faraday tensor $\boldsymbol{F}$. Furthermore, $F_{\mu\nu}=\partial_{\mu}A_{\nu}-\partial_{\nu}A_{\mu}$ for $A_{\mu}$ the one-form potential. Variation of expression \eqref{eq:bareaction} with respect to the metric leads to the Einstein equations,
\begin{equation} \label{eq:einstein}
    R_{\mu \nu} - \frac{1}{2} g_{\mu \nu} R = 8 \pi T_{\mu \nu},
\end{equation}
for stress-energy $T_{\mu}^{\nu} = - 2 \LEM'(F) F_{\mu \alpha}F^{\nu \alpha} + \tfrac{1}{2} \delta^{\nu}_{\mu} \LEM$ \cite{bronn23}. The variation of the action with respect to the potential $A_{\mu}$ yields   
\begin{eqnarray}
 {\nabla_{\mu}\left(  \frac{d\mathcal{L}_{\text{EM}}}{dF}\,F^{\mu\nu}\right)=0, }
\end{eqnarray}
{from which it follows that}
\begin{eqnarray}\label{eofM}
\partial_{\mu}\!\left(\sqrt{-g}\,\frac{d\mathcal{L}_{\text{EM}}}{dF}\,F^{\mu\nu}\right) = 0.    
\end{eqnarray}

To describe the electromagnetic sector more conveniently, it is useful to introduce an auxiliary antisymmetric tensor field \( \boldsymbol{P} \) defined by  
\begin{equation} \label{eq:ptensor}
    P_{\mu\nu} = \frac{d\mathcal{L}_{\text{EM}}}{dF}\, F_{\mu\nu}.
\end{equation}
With this definition, Eq.~(\ref{eofM}) takes the form of the “standard” Maxwell equation for $ P_{\mu\nu}$. Moreover, this relation suggests a structure reminiscent of a Legendre transformation, which motivates the definition of the Hamiltonian-like function  
\begin{equation} \label{eq:legendre}
\mathcal{H} = 2 F \frac{d\mathcal{L}_{\text{EM}}}{dF} - \mathcal{L}_{\text{EM}}.
\end{equation}
which can be thought of as a function of the rescaled invariant $P = \frac{1}{4}P_{\mu\nu}P^{\mu\nu}$. In this sense, the Lagrangian can be written
\begin{equation} \label{eq:lagrangian}
\LEM =  2 P \mathcal{H}_{P} - \mathcal{H} ,
\end{equation}
such that the Faraday tensor reads
\begin{equation} \label{eq:faraday}
F_{\mu\nu} =  \mathcal{H}_{P} P_{\mu\nu} ,
\end{equation}
where $\mathcal{H}_{P}=d \mathcal{H}/d P$. Using the above, the energy-momentum tensor is re-expressed as
\begin{equation} \label{eq:stresstensor}
T_{\mu\nu} = \frac{1}{4 \pi}\mathcal{H}_{P} P_{\mu\alpha}P_\nu^\alpha - \frac{1}{4 \pi} g_{\mu\nu} \left(2 P 
\mathcal{H}_{P} - \mathcal{H} \right).
\end{equation}
Note that the Maxwell case is recovered when $\mathcal{H}_P=1$, as expected.

The advantage of the above is that the Einstein equations \eqref{eq:einstein} simplify in such a way that one can easily ``invert'' them to deduce the nature of $\LEM$ from a given seed \eqref{eq:metric}. The details depend on whether one considers magnetically- or electrically-charged BHs, as this adjusts the non-zero components of the Faraday tensor. In the general case of a dyonic object but with the simplification that $\delta = 0$, for instance, one finds that the Faraday invariant is set by the transcendental relation
\begin{equation} \label{eq:generalfarad}
    F(r) = \frac{2}{r^4} \left\{ Q_{m}^2  - \frac{Q_{e}^2}{\left[\LEM'(F(r))\right]^2} \right\},
\end{equation}
for electric and magnetic charges $Q_{e}$ and $Q_{m}$, respectively, and where a prime denotes differentiation with respect to the argument. These charges are defined through 
\begin{equation} \label{eq:charges}
Q_{e} = r^2 \LEM'(F) F^{tr},  \qquad Q_{m} = F_{\theta \phi} \csc \theta,
\end{equation}
with all other components of $\boldsymbol{F}$ vanishing. From expression \eqref{eq:generalfarad}, an \emph{exact} solution to the equations of motion \eqref{eq:einstein} is obtained if one now chooses $\LEM$ such that 
\begin{equation} \label{eq:invert}
\mathcal{H}(P(r)) = -\frac{m'(r)}{r^2},
\end{equation}
where $m(r)$ is the mass function defined via 
\begin{equation} \label{eq:massfunc}
    N(r) = 1 - \frac{2m(r)}{r}.
\end{equation}
The value of the P-dual formalism is evidenced by the simplicity of equation \eqref{eq:invert}. For the remainder of this paper, we consider only the case of electrically-charged solutions and hence set $Q_{m} = 0$, dropping the subscript $e$ so that the symbol $Q$ denotes the electric charge. In particular, the electric field is found through
\begin{equation} \label{eq:electricfield}
    E(r) = \frac{Q_{e}}{r^2 \LEM'(F)}.
\end{equation}
An important feature of the NLED term within the denominator in expression \eqref{eq:electricfield} is that the electromagnetic energy of a point charge,
\begin{equation} \label{eq:energyintegral}
    U_{\rm em} \propto \int^{\infty}_{0} dr r^2 E(r)^2,
\end{equation}
may be finite \cite{Herdeiro:2019iwl}. NLED effects may thus not only resolve the singularity problem associated with the BH but the classical divergence of the Coulombic field ($E = Q/r^2$), for which $U_{\rm em} \to \infty$.

Given a choice of $N$, one need only solve equations \eqref{eq:generalfarad} and \eqref{eq:invert} for $F(r)$ and $\LEM$ (utilising \ref{eq:lagrangian}), respectively, to find an exact theory around the metric. A specific example of a well-motivated background is considered in Sec.~\ref{sec:examplesol}, though we now turn to the main question posed in this work: how does a general, regular BH respond to a non-minimal scalar coupling?

\section{Spontaneous scalarization} \label{sec:scalarization}

In coupling the scalar field to the dynamics, we consider the case
\begin{equation}\label{eq:scalaraction}
    \mathcal{S}=\int \text{d}^4x \sqrt{-g} \left[R-2\nabla_{\mu}\phi\nabla^{\mu}\phi-f(\phi)\LEM \right],
\end{equation}
where $\phi$ is a (real) scalar field, $\LEM$ is the Lagrangian associated to some NLED model and $f(\phi)$ is a coupling function that is responsible for triggering the scalarization provided certain conditions are met. It should be chosen such that the ``hairless'' solutions (i.e. with $\phi=0$) exist, which requires $f'(0)=0$. More generally, the function $f$ should satisfy certain identities to (i) prevent the emergence of ghosts (similar to constraints on symmetron-like theories, for instance \cite{koy16}) and (ii) permit the existence of scalarized solutions by satisfying the so-called Bekenstein identities \cite{bek72}. The latter essentially encapsulate virial relations required to ensure the existence of real-valued solutions for $\phi$, reading  \cite{Herdeiro:2018wub,Yu:2020rqi}
\begin{equation} \label{eq:bekenstein}
    f''(\phi) >0 , \qquad \phi f'(\phi) >0.
\end{equation}

By varying the action \eqref{eq:scalaraction} with respect to the scalar field, we arrive at the (in general nonlinear) equation
\begin{equation} \label{eq:kleing}
0 = \nabla_{\mu} \nabla^{\mu} \phi - \frac{f'(\phi) \LEM}{4}.
\end{equation}
In considering a small perturbation about the hairless solution, $\phi \to \delta \phi$,  one finds
\begin{equation}\label{eq:perturbation}
    0 = \left[ \nabla^{\mu} \nabla_{\mu} -\frac{f''(0) \LEM}{4}\right]\delta\phi + \mathcal{O}(\delta \phi^2),
\end{equation}
which one can identify as the Klein-Gordon equation with an effective mass
\begin{equation} \label{eq:effectivemass}
    \mu_{\rm eff}^2 = \frac{f''(0) \LEM}{4}.
\end{equation}
As such, if $f''(0) \LEM < 0$ the effective mass is \emph{imaginary} and a tachyonic instability can occur such that exponentially-growing modes lead to the generation of large-amplitude scalar hair (see Ref.~\cite{doneva24} for a review). A suitable choice, widely considered in the literature (see, e.g., Refs.~\cite{Herdeiro:2018wub,Yu:2020rqi,Herdeiro:2020htm,Carrasco:2024}) satisfying the Bekenstein conditions \eqref{eq:bekenstein}, is the function
\begin{equation} \label{eq:couplingfunc}
    f(\phi)=e^{-\alpha\phi^2},
\end{equation}
for some coupling constant $\alpha$. Noting that $\LEM \leq 0$ in most cases of interest (e.g., in the Maxwell limit we have $\LEM = F_{\mu \nu}F^{\mu \nu} \leq 0$) and $f''(0) = -2 \alpha$, it is clear that for $\alpha <0$ we have $\mu_{\rm eff}^2 < 0$ from expression \eqref{eq:effectivemass} and thus scalarization may occur.

The equations of motion can be derived from the action \eqref{eq:scalaraction} with the choice \eqref{eq:couplingfunc}. Alternatively, they can be obtained from the effective Lagrangian
\begin{eqnarray}\label{eq:effective-lagrangian}
\mathcal{L}_{\text{eff}} = e^{-\delta} m' 
- \frac{1}{2} e^{-\delta} r^{2} N \phi'^{2}
- r^{2} e^{-\delta - \alpha \phi^{2}} \mathcal{L}_{\text{EM}},
\end{eqnarray}
by varying with respect to the dynamical fields 
$m, \delta, \phi$, and $A_0 = V(r)$ for static, and  purely electric configurations. 

The Euler--Lagrange equation associated with $V$ is  
\begin{eqnarray}
\left(e^{\delta - \alpha \phi^{2}} r^{2} V'\right)' = 0,
\end{eqnarray}
which can be easily integrated as  
\begin{eqnarray}\label{V}
\frac{d\mathcal{L}_{\text{EM}}}{dF}\,V' = -\,\frac{Q}{e^{\delta - \alpha \phi^{2}} r^{2}}.
\end{eqnarray}
In the above expression, $Q$ is an integration constant interpreted as the electric charge. The remaining equations follow straightforwardly and read  
\begin{align} 
m'&=-r^2 \mathcal{H}e^{-\alpha\phi^2}+\frac{1}{2}r^2 N \left(\phi'\right)^2,\label{eq:eom1}\\
\delta'&=-r\left(\phi'\right)^2,\label{eq:eom2}\\
(r^2\phi' N e^{-\delta})'&=2\alpha r^2\phi e^{-\alpha\phi^2-\delta}\left(\mathcal{H}+\frac{1}{r^4}\mathcal{H}_{P}Q^2e^{2\alpha\phi^2}\right),\label{eq:eom3}
\end{align}
where we have used (\ref{V}) together with the definition \eqref{eq:lagrangian}.  

At this point, a few remarks are in order. First, note that when $\mathcal{H}_{P}=1$, the set of equations (\ref{V})--(\ref{eq:eom3}) reduces to the standard Maxwell case reported in \citet{Herdeiro:2018wub}. Secondly, Eqs.~(\ref{V})--(\ref{eq:eom3}) are generally valid for any $\mathcal{H}(P)$, provided the assumptions of staticity and the absence of magnetic monopoles hold. In this regard, for any given model $\mathcal{H}(P)$, one can insert it into the above expressions.   Next, note that the equations depend only on the radial coordinate. In fact, from (\ref{eq:ptensor}) we obtain  
\begin{eqnarray}
P=-\frac{1}{2}e^{-2\delta}\left(\frac{d\mathcal{L}_{\text{EM}}}{dF}V'\right)^{2},
\end{eqnarray}
which, using (\ref{V}), leads to  
\begin{eqnarray}\label{eq:P}
P=-\frac{Q^{2}}{2r^{4}}e^{2\alpha \phi^{2}}.   
\end{eqnarray}

Finally, note that because of the multiplicative coupling of the scalar and electromagnetic sectors in the action \eqref{eq:scalaraction}, the electric field is unaffected by the scalar field except through a rescaling of the metric functions. Before concluding this section, it is worth emphasizing that the methodology developed here is completely general and applies to any NLED model. This makes it particularly useful in situations where an explicit form of $ L_{EM}(F)$ cannot be obtained, as shown in \cite{Carrasco:2024,zhang24,zhang25}. An alternative approach was proposed in Ref.~\cite{Muniz:2024wiv}, although it is restricted to specific cases such as power--Maxwell models, for example.

\subsection{Domain of existence} \label{sec:domainofexistence}

Before attempting to solve the full system \eqref{eq:eom1}--\eqref{eq:eom3} together with the scalar equation \eqref{eq:kleing}, we examine the \emph{domain of existence}. This set corresponds to the valid range of metric and theory parameters such that a scalarized branch of solutions exists; identifying this set simplifies the subsequent numerical analysis (see Sec.~\ref{sec:numerics}).
In general, a scalar field perturbation over a static and spherically-symmetric background can be written
\begin{equation}\label{eq:decomposition}
    \delta \phi(t,r,\theta,\varphi)=\sum_{\ell,m} e^{i \omega_{\ell m} t}Y_{\ell,m}(\theta,\varphi)U_{\ell}(r),
\end{equation}
for spherical harmonics $Y_{\ell m}$, eigenfunctions $U_{\ell}$ (which do not depend on $m$ due to spherical symmetry), and eigenfrequencies $\omega_{\ell m}$. It is then straightforward to obtain the  equation of motion for the radial eignefunctions from \eqref{eq:perturbation} as 
{
\begin{equation}\label{eq:scalarpert0}
    -\frac{e^{2\delta}\omega^{2}_{lm}}{N}U_{\ell} = \frac{e^{\delta}}{r^2}\frac{d}{dr}\left(\frac{r^2 N}{e^{\delta}}\frac{dU_{\ell}}{dr}\right)-\left[\frac{\ell(\ell+1)}{r^2}+\mu_{\rm eff}^2\right]U_{\ell}.
\end{equation}
Now, a necessary condition for the tachyonic instability to occur is that frequency becomes imaginary. 
Thus, the boundary between positive and negative values of $\omega^{2}_{\ell m}$ is defined by the condition $\omega^{2}_{\ell m}=0$, which leads to
}
\begin{equation}\label{eq:scalarpert}
    0 = \frac{e^{\delta}}{r^2}\frac{d}{dr}\left(\frac{r^2 N}{e^{\delta}}\frac{dU_{\ell}}{dr}\right)-\left[\frac{\ell(\ell+1)}{r^2}+\mu_{\rm eff}^2\right]U_{\ell}.
\end{equation}
If $\mu_{\rm eff} < 0$, the tachyonic instability can occur as described in Sec.~\ref{sec:scalarization}. The solution of equation \eqref{eq:scalarpert} together with Dirichlet boundary conditions to fix $\omega_{\ell m}$ defines the lower bound -- the so-called \emph{existence line} -- for scalarized solutions: a given set of parameters sets the minimal, necessary conditions for exponentially-growing modes. In the simple case we are considering, we need only consider the fundamental $\ell = 0$ eigenfunction ($U_0$), as this defines the strongest modes, so that the task of obtaining the existence line reduces to identifying the zeros of $U_0$ at infinity ($r \to \infty$).

Aside from the existence line, there similarly exists a \emph{critical set} -- setting an upper bound for the charge for given $\alpha$ such that a scalarized branch exists -- defined by a vanishing horizon area (i.e., to ensure a BH rather than naked singularity).

\subsection{Numerical methods} \label{sec:numerics}

The system (\ref{eq:eom1})–(\ref{eq:eom3}) cannot be solved analytically in general; therefore, we employ the shooting method (see, for instance, Refs.~\cite{Herdeiro:2018wub,Carrasco:2024} and references therein for applications). This method consists of an iterative procedure in which initial trial values for the unknowns are adjusted until the boundary conditions are satisfied. In our model, the functions ${m,\delta,\phi}$ must be determined for any $r \in (r_H, \infty)$. Since we require asymptotically flat solutions, the boundary conditions at infinity are $m \to M$, $\delta \to 0$, and $\phi \to 0$. However, the exact values of $\delta$ and $\phi$ at the horizon $r_H$ remain undetermined. To reduce the number of free parameters, we expand the functions near the horizon as
\begin{eqnarray}
m(r) &=& \frac{r_H}{2} + m_1 (r - r_H) + \cdots,\\
\delta(r) &=& \delta_0 + \delta_1 (r - r_H) + \cdots,\\
\phi(r) &=& \phi_0 + \phi_1 (r - r_H) + \cdots.
\end{eqnarray}
Replacing this expansions in the equations of motion, we arrive at
\begin{eqnarray}
m_1&=& -e^{-\alpha\phi_0^2} r_H^2  \mathcal{H}_0\\
\delta_1&=&-\phi_1^2r_H\\
\phi_1&=&\frac{2 \alpha  \phi _0 \left(Q^2 e^{2 \alpha  \phi _0^2} \mathcal{H}_{P0}+r_H^4 \mathcal{H}_0\right)}{r_H^3 \left(2 r_H^2 \mathcal{H}_0+e^{\alpha  \phi _0^2}\right)},
\end{eqnarray}
where $\mathcal{H}_0=\mathcal{H}\left(r_H,\phi _0\right)$ and $\mathcal{H}_{P0}=\mathcal{H}_{P}\left(r_H,\phi _0\right)$. 

From the above expansions, we identify our unknowns as $\delta(r_H) = \delta_0$ and $\phi(r_H) = \phi_0$. Furthermore, note that our system remains invariant under the transformation $\delta \rightarrow \delta + \tilde{\delta}$, where $\tilde{\delta}$ is a constant. Therefore, we can initially set $\delta_0 = 0$ and later use this symmetry to recover the physical solutions. Consequently, the number of unknown parameters is reduced to a single one, $\phi_0$, which serves as the shooting parameter.  Fixing $\{r_H, Q\}$, we choose an initial guess for the shooting parameter, integrate the equations of motion, and repeat the process, updating the value of $\phi_0$ until $\phi$ vanishes at (numerical) spatial infinity.

A useful diagnostic for determining the validity of a numerical solution is a generalised virial identity. This criterion is a necessary condition for the existence of soliton-like configurations in field theories (e.g. \cite{herd21}), placing constraints on the effective potential such that the scalar field remains real. In our theory of interest \eqref{eq:scalaraction}, the virial identity reads
\begin{eqnarray} \label{eq:virial}
&&\int\limits_{r_{H}}^{\infty}dr r^{2}e^{-\delta-\alpha\phi^2}\left[ \left(3-\frac{2r_{H}}{r}\right)\mathcal{H}-4P\mathcal{H}_{P}\left(1-\frac{r_{H}}{r}\right)\right]\nonumber\\
&&\hspace{40pt}= \int\limits_{r_{H}}^{\infty}dr\frac{e^{-\delta}r^2 \phi'}{2}\left[1+\frac{2r_{H}}{r}\left(\frac{m}{r}-1\right)\right]    
\end{eqnarray}
Both sides of expression \eqref{eq:virial} are checked for each numerical solution, and if the mismatch lies below a tolerance (set as $10^{-6}$ in appropriate, dimensionless units) the result is considered acceptable.

\section{Worked example: Balart-Vagenas black hole} \label{sec:examplesol}

For the sake of providing a worked example, we carry out the above process for a particular exact solution. Consider the metric introduced by \citet{bv14} (see their equation 25)
\begin{equation} \label{eq:nfn}
N(r) = 1 - \frac{2 M}{r} \left\{1-{\left[1+\left(\frac{2 M r}{Q^2}\right)^3\right]^{-1/3}}\right\},
\end{equation}
for mass $M$, charge $Q$, and $\delta(r) = 0$. The relevant ``Maxwell'' equations \eqref{eofM} are satisfied for electric field given by
\begin{equation}
    E(r) = \frac{Q}{r^2} \left(1 + \frac{Q^6}{8 M^3 r^3} \right)^{-7/3},
\end{equation}
which clearly has the Coulomb limit at large radii. In particular, the metric defined by \eqref{eq:nfn} is regular with all curvature invariants being bounded as $r\to0$ for $|Q| >0$ \cite{bv14} and has finite electromagnetic energy \eqref{eq:energyintegral}. The solution possesses an event horizon provided that $Q \leq 1.0257M$. For this metric, the P-dual formalism detailed in Sec.~\ref{sec:inverse} is used to introduce the function $\mathcal{H}$ appearing within the Lagrangian \eqref{eq:lagrangian} as 
\begin{equation} \label{eq:someh}
\mathcal{H}(P) = \frac{P}{\left[1 + \Upsilon  \left(-P\right)^{3/4}\right]^{4/3}},
\end{equation}
where $\Upsilon$ is given by
\begin{equation} \label{eq:uspilon.}
\Upsilon = \left[\frac{Q^{3/2}}{2^{3/4} M}\right]^3.
 \end{equation}

The above fully characterise the \emph{unscalarized} solution. For a discussion on some astrophysical tests of the solution, we refer the reader to Ref.~\cite{zhang22}.

\begin{figure}
    \centering
    \includegraphics[width=0.5\textwidth]{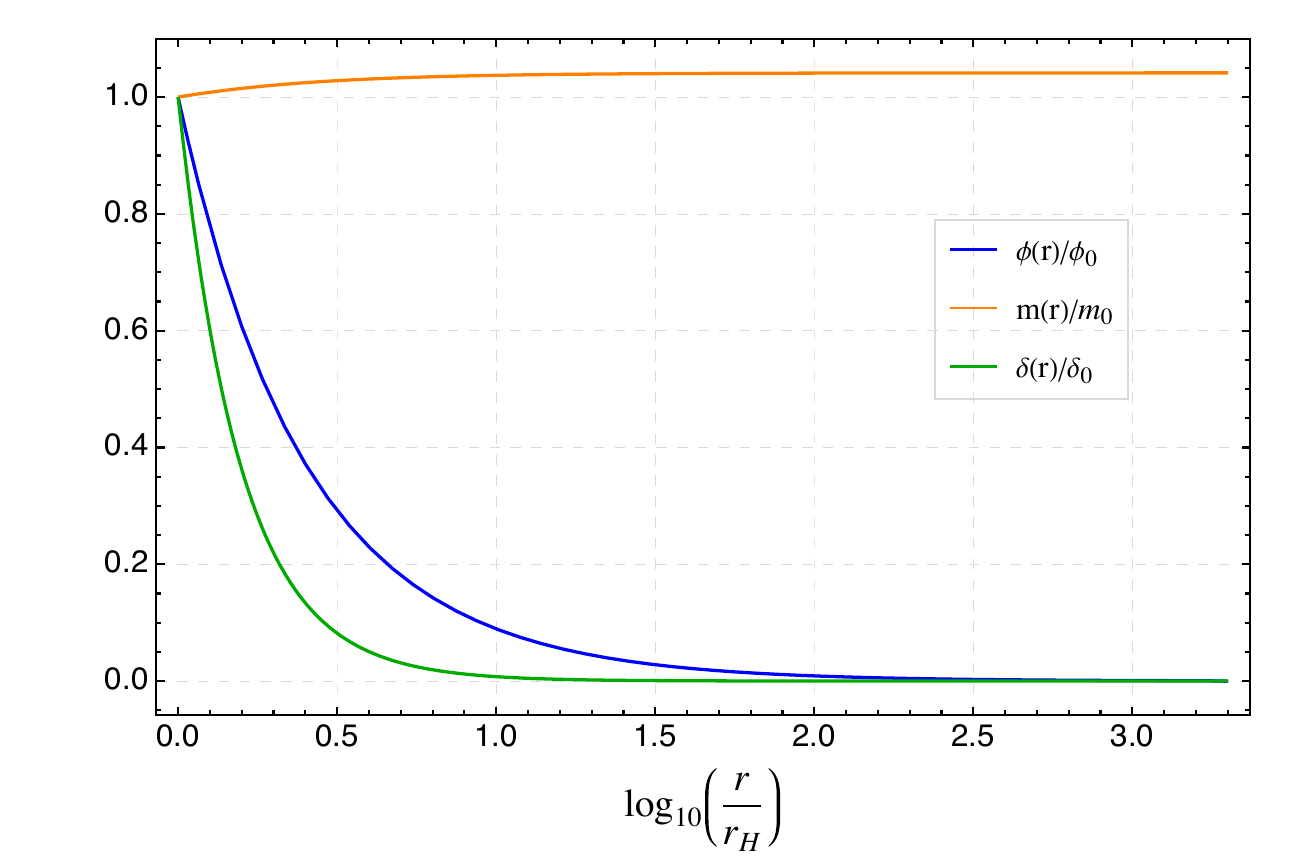}
    \caption{Normalized scalar field, $\phi$, mass function, $m$, and lapse, $\delta$ (see figure legends) as a function of the radius for $Q=0.1$, $M=0.9$, 
    $\alpha=-30$ and $r_H=0.3$.} 
    \label{fig:phi_vs_r}
\end{figure}

In what follows, we proceed with the numerical integration of the equations of motion following the strategy described in the preceding section. To this end, we replace the Hamiltonian $\mathcal{H}$, given by (\ref{eq:someh}), and its derivative $\mathcal{H}_P$. We then express the equations solely in terms of the radial coordinate by using (\ref{eq:P}). The shooting parameter is $\phi_0$. A representative solution is shown in Fig.~\ref{fig:phi_vs_r} for the parameters $Q=0.1$, $M=0.9$, $\alpha=-30$, and $r_H=0.3$. The solution behaves as expected: the scalar field vanishes asymptotically, while the mass increases monotonically with the radius. 

\begin{figure}[h!]
    \centering
    \includegraphics[width=0.5\textwidth]{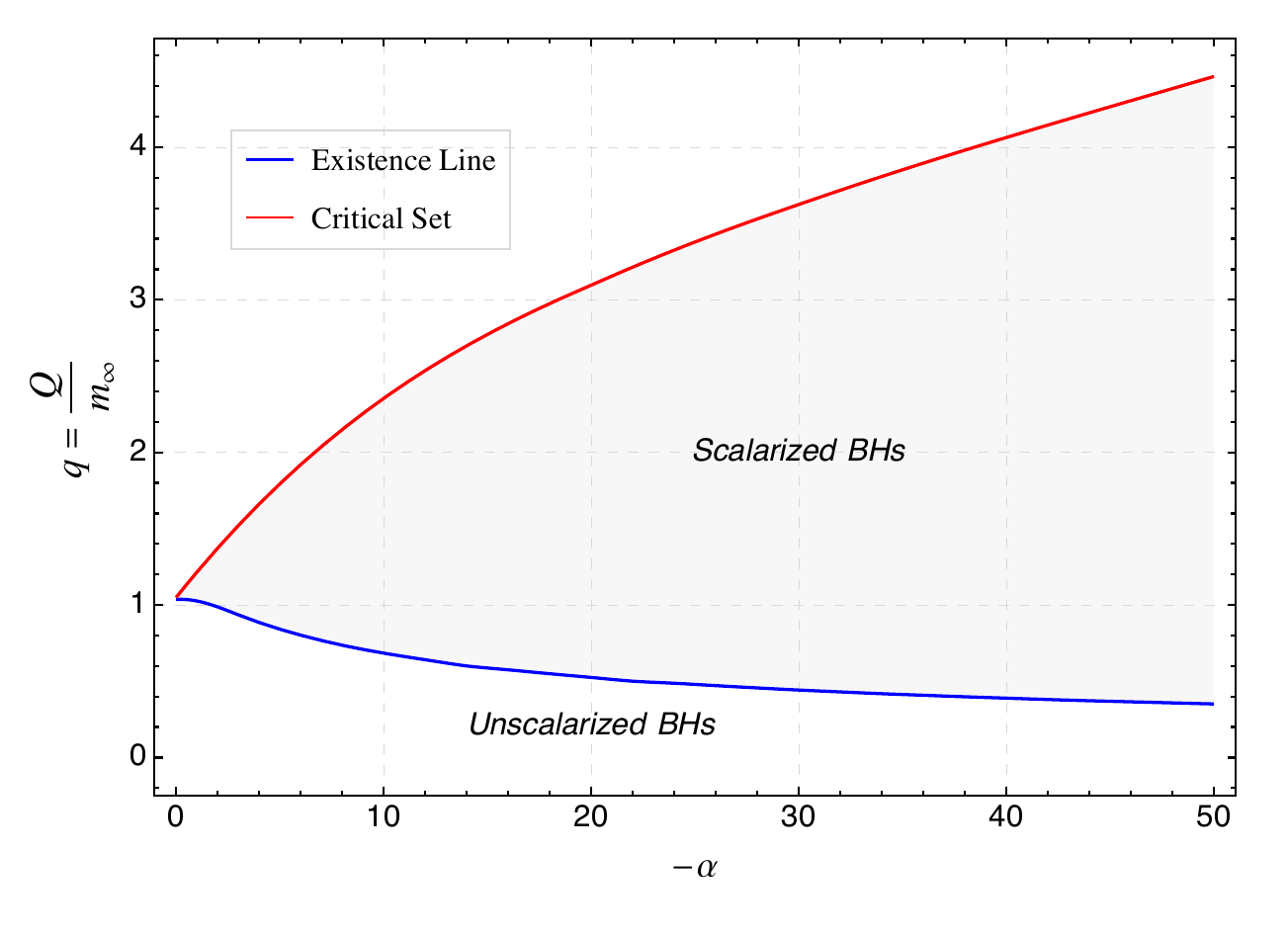}
    \caption{Domain of existence for spherical scalarized regular black holes as a function of $-\alpha$ and $q=\frac{Q}{m_{\infty}}$, where $m_\infty=m(r\to\infty)$.}
    \label{fig:m_vs_r}
\end{figure}

\begin{figure} [h!]
    \centering
    \includegraphics[width=1\linewidth]{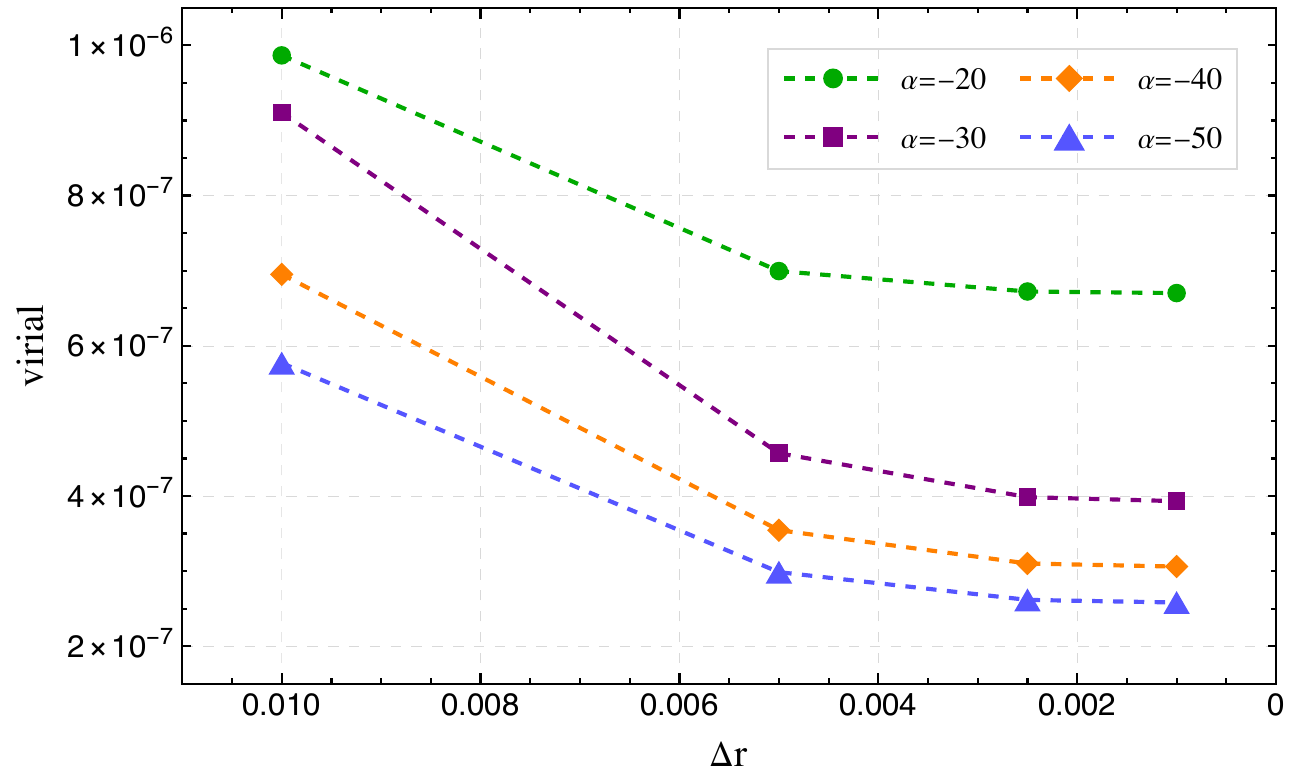}
    \caption{Virial identity as a function of the numerical step size. The results show that the virial identity converges to zero in the limit of vanishing step size.}
    \label{virialplot}
\end{figure}

The domain of existence is obtained through numerical iteration by fixing $\alpha$ and $Q$, and varying $r_H$. For each $r_H$, the equations of motion are solved, and the initial guess for $\phi_0$ is taken from a neighboring solution. Each $\alpha$-branch starts at the existence line (obtained by integrating (\ref{eq:scalarpert})) and ends at the \textit{critical set}, defined by a vanishing horizon area. The scalarized solutions exhibit a virial mismatch on the order of $10^{-6}$. The domain of existence is shown in Fig.~\ref{fig:m_vs_r}. {At this point it is important to highlight that we have explicitly checked 
the numerical convergence of our solutions. In particular, we repeated the 
integration by using different 
resolutions (varying the radial step size) and verified that the deviation from  the virial identity systematically decreases as the step size is reduced as shown in Fig. \ref{virialplot}. For instance, choosing $\Delta r \leq 0.006$ (again in dimensionless units) shows that the virial identity is numerically satisfied within our specified tolerance of $10^{-6}$ for any values of $-\alpha \leq 50$. Moreover, the monotonic reduction of the mismatch as a function of decreasing $\Delta r$ shows the scheme converges with increasing resolution, as expected.
}

\begin{figure} [h!]
    \centering
    \includegraphics[width=1\linewidth]{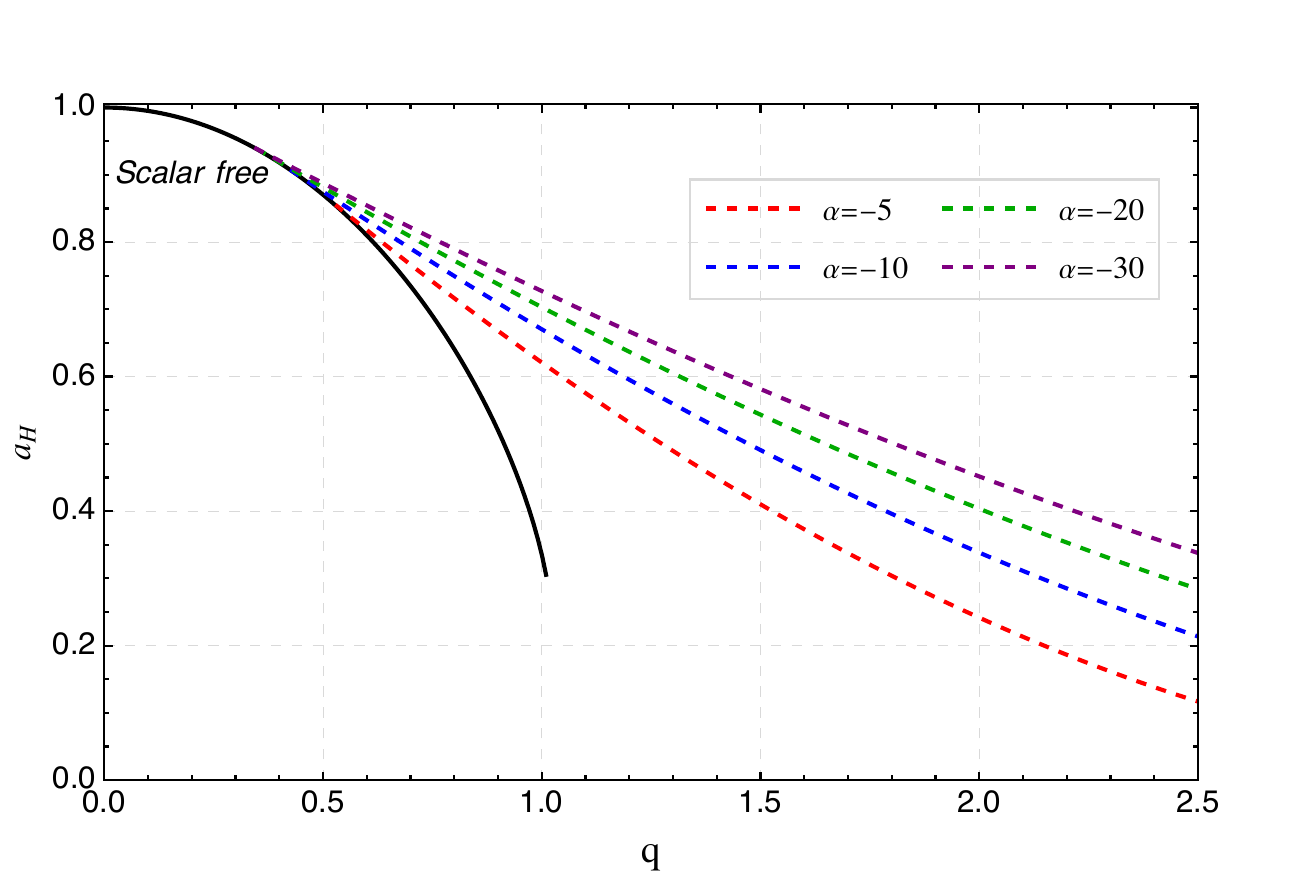}
    \caption{Normalized area, $a_H=\frac{A_H}{4\pi m_\infty}$, as a function of $q$, where $A_H=4\pi r_H^2$. The entropy is proportional to the area in the theories under consideration.}
    \label{fig:area_vs_q}
\end{figure}

It is worth noting that there exists a region of non-uniqueness within the domain of existence where scalar-free and scalarized BHs coexist, as defined by the values of $q = Q/m_\infty$ for which the metric $N$ possesses real roots. In this region, the scalarized solutions are entropically preferred, as they maximize the entropy (or, equivalently, the horizon area $A_H$), as shown in Fig.~\ref{fig:area_vs_q} for the specific values indicated in the legend. It is clear that the entropically-preferred solutions correspond to those for which the coupling parameter $|\alpha|$ increases.

\section{Astrophysical observables} \label{sec:astro}

\subsection{Shadows} \label{sec:shadows}

As a first simple application, we consider the deflection of photons in the spacetime described by the scalarized and unscalarized branches in the context of geometric optics. Photon trajectories are relevant, for instance, in the determination of BH shadows \cite{sch92} to compare models with astrophysical observations made by the Event Horizon Telescope (EHT; \cite{eht1}). While astrophysical holes probed by EHT observations are expected to rotate rapidly and thus we are unable to make direct comparisons with observational data in this paper, we can explore qualitative trends to pave the way for future examinations of stationary spacetimes in theories described by the action \eqref{eq:scalaraction}.

Starting from the particle Euler-Lagrange equations,
\begin{equation} \label{eq:eulerlagrange}
0 = \frac{d}{d \lambda} \left( \frac{\partial \mathscr{L}}{\partial \dot{x}^{\mu}} \right) -  \frac{\partial \mathscr{L}}{\partial x^{\mu}},
\end{equation}
where $\mathscr{L} = \tfrac{1}{2}g_{\mu \nu} \dot{x}^{\mu} \dot{x}^{\nu}$ for affine parameter $\lambda$ and particle 4-momenta $\dot{\boldsymbol{x}}$, the \emph{critical impact parameter} for the metric \eqref{eq:metric} is given by \cite{per22}
\begin{equation} \label{eq:genbcrit}
b_{\rm cr}^2 = \left[ \frac{r^2}{N(r) e^{-2 \delta(r)}} \right]_{{r = r_{\rm ph}}},
\end{equation}
where $r_{\rm ph}$ denotes the radius of the photon sphere. Physically, $b_{\rm cr}$ denotes the impact parameter separating  flyby and captured orbits of incoming light rays. For a static, spherically-symmetric BH, it delimits the radius of the ``shadow'' seen by a distant observer. The photon sphere radius (or radii), $r_{\rm ph}$, is found as a solution to the equation
\begin{equation} \label{eq:rphoton}
    r_{\rm ph} = 3 m(r_{\rm ph}).
\end{equation}

\begin{figure}
    \centering
    \includegraphics[width=1\linewidth]{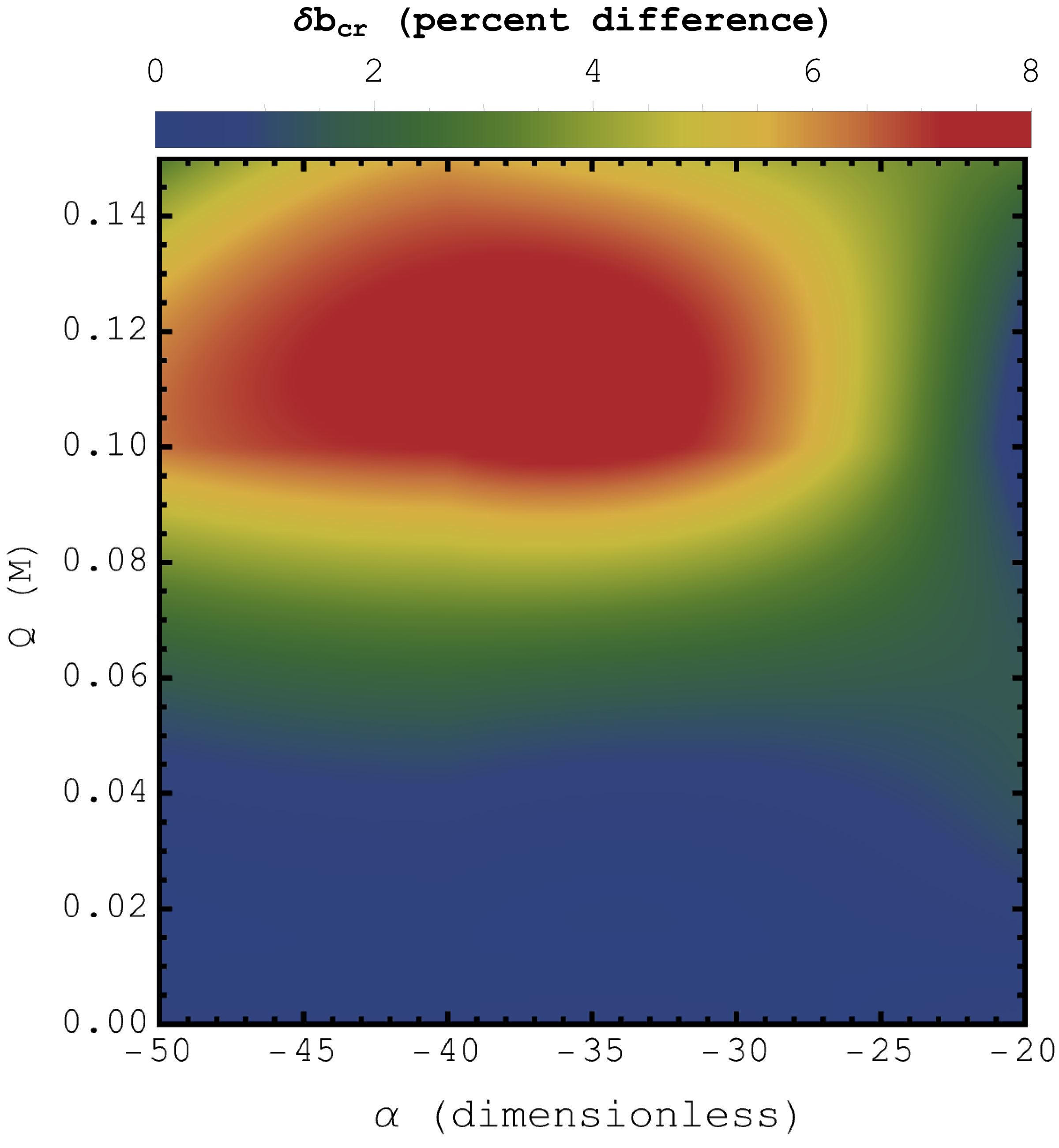}
    \caption{Relative differences between the critical impact parameter for scalarized solutions relative to unscalarized  ones, $\delta b_{\rm cr}$, as a function of charge, $Q$, and coupling constant, $\alpha$. Redder shades indicate greater departures.}
    \label{fig:deltab}
\end{figure}

Figure~\ref{fig:deltab} depicts how the critical radius, from expression \eqref{eq:genbcrit}, varies for a range $0 \leq Q \leq 0.15$ and $20 \leq -\alpha \leq 50$ between the (numerically-determined) scalarized and unscalarized solutions while keeping the total mass, $m_\infty$, fixed between each grid point. The figure is produced by using spacings $Q$ and $\alpha$ of 0.05 and 10, respectively, and using a quadratic interpolation. To facilitate a comparison, we have defined $\delta b_{\rm cr} = b^{\rm scalarized}_{\rm cr} - b^{\rm unscalarized}_{\rm cr}$ as a measure for the relative departure. In astrophysical applications, we expect that charge is comparatively small as a negatively (positively) charged BH will preferentially capture protons (electrons) and thus tend to neutralize over cosmological timescales. In such cases, we see that the change to the shadow radius is in the neighbourhood of zero: only once $\alpha \lesssim -30$ and $Q \gtrsim 0.05$ do changes at the percent-level appear. At the most extreme we find a change of $\approx 8\%$ for the largest values of $Q \sim 0.15$, though with the interesting feature that the relative difference is not monotonic with respect to $\alpha$. This occurs because scalarized branches tend to have larger mass (see, e.g., Fig.~\ref{fig:phi_vs_r}), which enlarges the photon sphere radius \eqref{eq:rphoton}, though the growth rate peaks at around $\alpha \approx -40$. This means that while both the numerator and denominators of equation \eqref{eq:genbcrit} increase monotonically, the relative ratio does not. Either way, given these relatively small changes it is unlikely that changes to the shadow shape will be discernible with near-future instruments between the two different solution branches.

\subsection{Quasi-normal modes} \label{sec:qnms}

The behaviour of scalar perturbations for the \emph{unscalarized} solutions was examined in Sec.~\ref{sec:domainofexistence} to determine the existence line. However, it is also of interest to study the scalar QNMs of the \emph{scalarized} solutions to see how they shift. Indeed, cases that are strongly scalarized (e.g., for $\alpha = -50$; see Fig.~\ref{sec:domainofexistence}) may lead to non-negligible departures in the predicted eigenfrequencies which could theoretically be detected by the ringdown signal of a newborn object. The recent event GW250114 with its unprecedented signal-to-noise ratio of $\approx 80$ has allowed for the some of the strongest tests of the Kerr hypothesis to date \cite{ligo25}. The analysis revealed an agreement with the Kerr spectrum to within $\sim$~tens of percent for the leading-order modes. As such, one may theoretically aim to place constraints on $\alpha$ by ensuring that the resultant spectrum does not depart by more than this number. 

Introducing scalar perturbations of the form \eqref{eq:decomposition} with $U_{\ell}(r) \to \Psi_{\ell}(r)/r$ and introducing the tortoise coordinate,
\begin{equation}
    r_{\star} = \int dr e^{-\delta} N(r),
\end{equation}
equation \eqref{eq:scalarpert} can be re-written in the Schr{\"o}dinger-like form
\begin{equation} \label{eq:mastereqn}
0 =    \frac{d^2 \Psi_{\ell}}{d r_{\star}^2} + \left[ \omega^2 - V_{\ell}\left(r_{\star}\right) \right],
\end{equation}
where, in the original coordinates, we have
\begin{equation} \label{eq:effectivepotential}
    V_{\ell}(r) =  \frac{N(r)}{e^{2 \delta(r)}} \left\{ \frac{\ell (\ell +1)}{r^2} + \frac{1}{r} \frac{d}{dr} \left[ e^{-\delta(r)} N(r) \right] \right\}.
\end{equation}
We determine the eigenvalues of equation \eqref{eq:mastereqn} satisfying the usual boundary conditions (i.e., purely ingoing waves at the horizon and purely outgoing ones at infinity) using the third-order Wentzel-Kramers-Brillouin (WKB) scheme introduced by Iyer and Will \cite{sai86,sai87}. While the use of a higher-order scheme would produce more accurate results for low values of $\ell$ away from geometric optics, this is sufficient for our purposes (see Ref.~\cite{ks99} for a review). In terms of the peak value for the potential \eqref{eq:effectivepotential}, denoted $r_{0}$ here with $V_{\ell}$ evaluated there as $V_{0}$, the complex eigenfrequencies $\omega$ are found through
\begin{equation} \label{eq:wkb}
\begin{aligned}
    \omega^2 =& \left[ V_{0} + (-2 V_{0}'')^{1/2} \Lambda_{2} \right] \\
    &- i \left(n+\frac{1}{2}\right) \left(-2 V_{0}''\right)^{1/2}\left(1 + \Lambda_{3}\right),
    \end{aligned}
\end{equation}
where the $\Lambda_{i}$ are higher-order WKB factors found, for instance, as expressions (1.3) in Ref.~\cite{sai87}. 

\begin{figure}[ht!]
    \centering
    \includegraphics[width=0.487\textwidth]{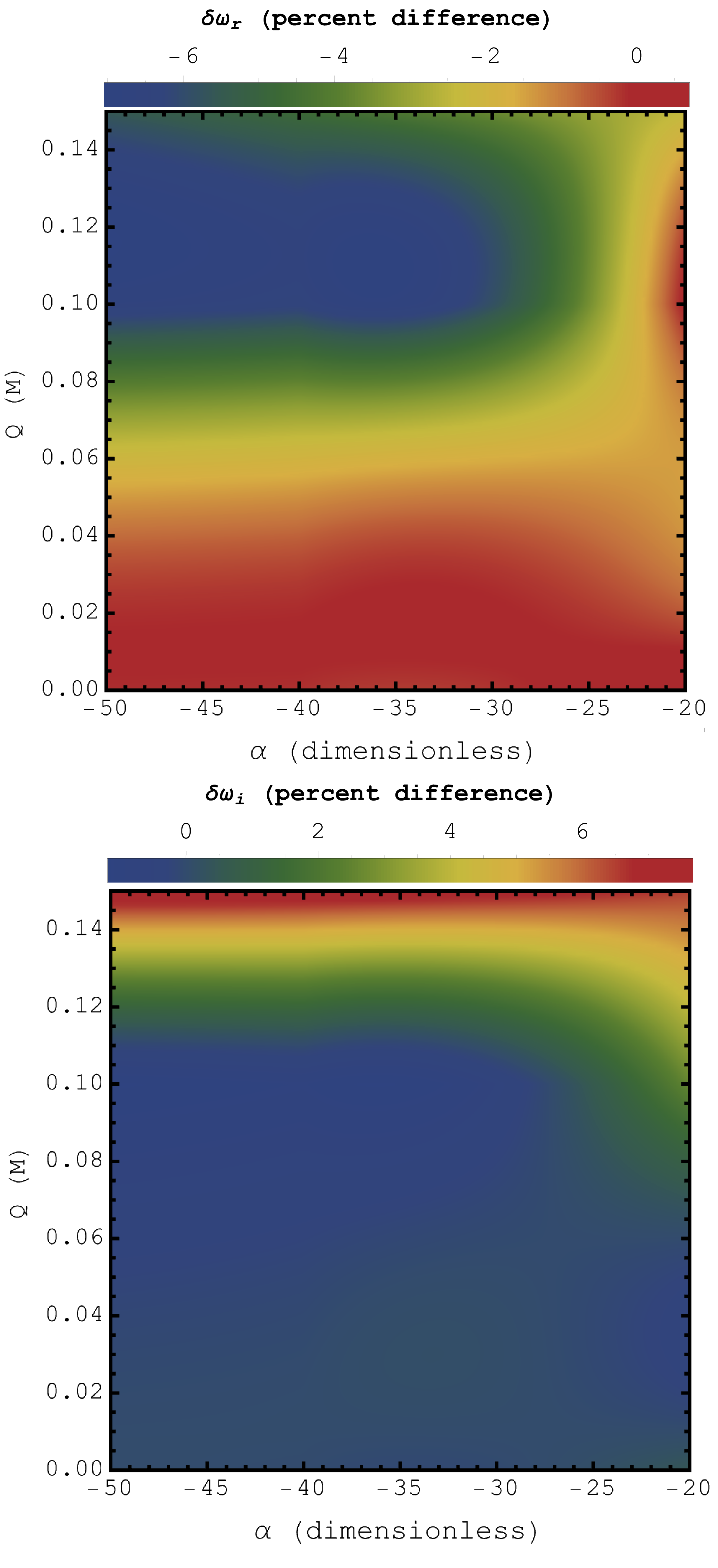}
    \caption{Similar to Fig.~\ref{fig:deltab} but showing changes to the real (top) and imaginary (bottom) parts of the fundamental, $\ell=2$ scalar QNMs computed using a 3rd-order WKB scheme.}
    \label{fig:deltaomega}
\end{figure}

In a presentation similar to Fig.~\ref{fig:deltab}, Fig.~\ref{fig:deltaomega} shows differences in the real (top) and imaginary (bottom) values of the fundamental ($n=0$), $\ell=2$ QNMs computed using expression \eqref{eq:wkb}---the dominant mode. We see that, in general, the real components tend to decrease while the imaginary ones increase. The latter implies that the modes are damped faster, as expected since enhancements to the scalar sector allow for more energy to be carried away as a function of time. Overall, changes to the real eigenfrequencies follow a pattern similar to that of the shadow radius as a function of both $Q$ and $\alpha$, which follows from the fact that the effective potential \eqref{eq:effectivepotential} is proportional to $N e^{-2\delta}$, which is precisely the same potential featuring in the denominator of expression \eqref{eq:genbcrit} for the critical impact parameter. Overall, changes are at a similar level to that of the shadow: for example, for $\alpha \approx -50$ and $Q \approx -0.15 M$ we have a $\approx 7\%$ change in both the real and imaginary components. Although we only consider static solutions here and scalar QNMs, such changes are comfortably within the constraints implied by GW250114 described above \cite{ligo25}. We may conclude that even for comparatively large values of the charge and scalar coupling, such solutions cannot be immediately ruled out by current gravitational-wave detections (at least at the level of ringdown). Future tests will allow for more stringent constraints.

\section{Conclusions} \label{sec:conclusions}

In this work we have presented a general and practical framework to study spontaneous scalarization of BHs sourced by nonlinear electrodynamics. The approach starts from an arbitrary static, spherically symmetric seed metric and uses the P–dual formalism to reconstruct the electromagnetic dynamics through $\mathcal{H}(P)$, guaranteeing an exact solution of the Einstein equations in the unscalarized limit. A non-minimal, multiplicative coupling between a real scalar and the nonlinear electrodynamics sector, $f(\phi)\LEM$ with $f(\phi)=e^{-\alpha\phi^2}$, triggers scalarization whenever $\mu_{\rm eff}^2=\tfrac14 f''(0)\LEM<0$. The full system is then solved with a horizon-to-infinity shooting scheme, and validated by a generalized virial identity (satisfied to within one part in $\sim 10^{6}$).

We mapped the domain of existence of scalarized branches by combining: (i) the \emph{existence line}, obtained from the linear perturbation equation for the $\ell=0$ mode, and (ii) a \emph{critical set} defined by a vanishing horizon area. Within this domain there is a region of non-uniqueness where scalar-free and scalarized solutions coexist; there, the scalarized configurations are entropically preferred (larger $A_H$ for the same conserved charges). Applying the method to the regular Balart–Vagenas solution, we found explicit, scalarized solutions and quantified their observational imprints. The predicted deviations in the shadow size and in the fundamental scalar quasi–normal frequencies are typically at the percent level (up to $\sim 8\%$ in the explored range of $\alpha$ and $Q$), indicating compatibility with current EHT and gravitational–wave bounds.

While we have explored shadows (Sec.~\ref{sec:shadows}) and QNMs (Sec.~\ref{sec:qnms}) and found that there are only small changes between the scalarized and unscalarized branches, the presence of astrophysical charge could be detected in other ways. For example, plasma circling around an accretion disc will be affected, with charged particles and neutrals experiencing different forces (i.e., only the former succumb to Coulomb repulsion). It would be worthwhile to study accretion dynamics in the spacetimes constructed here, as the presence of non-negligible charge will likely have observable effects at X-ray wavelengths (see, e.g., Ref.~\cite{reh24}). 

Beyond that which we have explored here, the coupling of general relativity to nonlinear electrodynamics likely has important consequences for the highly magnetised class of neutron stars known as magnetars. Such objects possess field strengths reaching $\sim 10^{15}$~G, where Maxwellian electrodynamics is expected to break down. Their hydromagnetic structure is sensitive to the particulars of the electromagnetic sector at such high strengths \cite{sp25} and, if the GR sector is also altered, further nontrivial features may reveal themselves (e.g., changes in cooling tracks, long-term magnetic evolution, or dynamo activity). It would be interesting to explore their predicted X-ray polarization properties in the theories considered here, which can be sensitively probed with future instruments (such as the planned enhanced X-ray Timing and Polarimetry mission \cite{ge25}).

The framework presented here is broadly applicable to any static seed geometry supported by nonlinear electrodynamics and readily extends to other coupling choices $f(\phi)$, provided the Bekenstein identities and the relevant energy conditions hold. Natural next steps include: (i) stability analyses beyond the virial test (time-domain evolutions and full mode spectra), (ii) extensions to dyonic and rotating spacetimes, (iii) dynamical formation and transition between branches, (iv) refined constraints from shadows, ringdowns, and strong-field lensing, and (v) other models inspired in alternative approaches \cite{Arias:2022jax,Bargueno:2022yob}. These avenues will sharpen the viability of scalarized, nonlinear electrodynamics-supported BHs as signatures of beyond–GR physics in the strong-field regime (including, perhaps, for those with nontrivial topology \cite{rin19,cap24} or non-Minkowski cores \cite{kh15}).

\begin{acknowledgments}
{We thank the anonymous referee for their helpful comments.}
E. C. and P.B acknowledge financial support from the Generalitat Valenciana through PROMETEO PROJECT CIPROM/2022/13. E. C. is funded by the Beatriz Galindo contract BG23/00163 (Spain).
M. C. H. thanks ICTP (Italy) where this work was initiated.
AGS acknowledges funding from the European Union's Horizon MSCA-2022 research and innovation programme ``EinsteinWaves'' under grant agreement No. 101131233.
\end{acknowledgments}

%

\end{document}